\documentclass[12pt,preprint]{aastex}

\newcommand{\kms}{\ensuremath{\mathrm{km\ s}^{-1}}}
\newcommand{\Ro}{\mbox{R$_{\odot}$}}
\newcommand{\Mo}{\mbox{M$_{\odot}$}~}
\newcommand{\beq}{\begin{equation}}
\newcommand{\eeq}{\end{equation}}
\newcommand{\Mdot}{\dot{M}~}

\shorttitle{Evolution of the planetary nebula NGC 1360}
\shortauthors{Garc\'{\i}a-D\'{\i}az et al.}

\begin{document}


\title{The planetary nebula NGC 1360, a test case of  magnetic collimation and evolution after the fast wind\\
    }


\author{M. T. Garc\'{\i}a-D\'{\i}az \altaffilmark{1},  J. A. L\'opez\altaffilmark{1}, G. Garc\'{\i}a-Segura,\\
 M. G. Richer \altaffilmark{1} \& W. Steffen\altaffilmark{1}}
\affil{Instituto de Astronom\'{\i}a, Universidad Nacional Aut\'onoma de M\'exico, Campus Ensenada,
    Ensenada, Baja California, 22800, M\'exico}







\begin{abstract}
The central star of this nebula has an observed intense magnetic field and the fast wind is no longer present, indicating that a back flow process has probably developed. 
Long-slit, spatially resolved echelle spectra have been obtained across the main body 
of NGC 1360 and over its system of bipolar jets. Deep images of the knotty structures of the
jets have also been obtained. The data allow a detailed study of the structure and kinematics
of this object and the results are modeled considering the effects of a magnetic collimation process in the development of the nebula and then switching off the fast stellar wind to follow its evolution to its current state. The model is able to successfully reproduce many key features of NGC 1360 under these premises.

\end{abstract}


\section{Introduction}
NGC 1360 is an interesting planetary nebula for several reasons. Firstly, this is one of the few planetary nebulae where an intense stellar magnetic field, of the order of kilogauss, has been detected in its core \citep{jor05}. Secondly, the nebula shows an elongated and nearly featureless morphology with no apparent bright rim, sharp inner boundary or central cavity, indicating the current absence of a significant stellar wind, this is further confirmed by archive IUE spectra of the core of NGC 1360 that reveal the absence of P Cygni-type line profiles. Furthermore, the nebula has a system of fast-expanding bipolar jets outside the main nebular body. 

Although NGC 1360 had been classified by \citet{men77} as containing a close binary nucleus, \citet{weh79} were unable to confirm such claim after a detailed radial velocity study of the central star. More recently, \citet{gol04} have described NGC 1360 as a thick shell without a sharp inner edge that indicates a lack of ongoing compression by a fast stellar wind. They also have brought attention to the presence of the bipolar jets and studied the kinematics of the shell and the northern jet. The central region of NGC 1360 shows no
[N~II] $\lambda$~6548, 6584 emission lines, these lines are only detected from the strings of knots that
form the system of bipolar collimated outflows and from an isolated knot located on the 
northern edge of the nebula in our data. 
Since collimated, bipolar outflows in planetary nebulae are thought to originate either from the action of binary nuclei  or  magnetic fields and given the lack of evidence of the former and the high significance of the observations of the latter in this case, NGC 1360 represents a unique case to test the possible effects of a magnetic collimation process in the development of this object. Furthermore, recently 
\citet{gar06} have started exploring the expected structure of planetary nebulae once the fast wind has faded, and NGC 1360 represents also an opportunity to examine the characteristics of this stage. 

Relevant data for NGC 1360 are its high electron temperature T$_e$ [O~III] = 16500 K   and low electron density N$_e$ $\lesssim$ 100 cm$^{-3}$ \citep{kal90}; high effective temperature
T$_{eff}$ = 97 000 K, log L = 3.64 L$_\odot$, M = 0.65 M$_\odot$, main sequence mass M$_{ms}$= 2.7 M$_\odot$ \citep{tra05, jor05}.  A distance estimate determined from  {\it Hipparcos} data \citep{ack98} d =${350} {\pm} {1000 \atop 180}$ pc is adopted here.

\section{Observations and results}
The long-slit observations were obtained with the Manchester Echelle Spectrometer (MES -- SPM) 
\citep{mea03} combined with the f/7.9 focus of the 2.1m San Pedro M\'artir UNAM telescope on 2004, October 29 and 30. This echelle spectrometer has no  cross-dispersion. For the present observations a filter of 90 \AA ~bandwidth was used to isolate the 87$^{th}$ order containing the H$\alpha$ and [N~II] nebular emission lines. A SITE CCD with $1024 \times 1024$ square pixels, each with 24 $\mu$m sides, was the detector. Two times binning was employed in both the spatial and spectral directions. Consequently 512 increments, each $0\farcs624$ long gave a projected slit length of $5\farcm32$ on the sky. "Seeing" varied between 1 - 2 arcsec during the observations. The slit was 150 $\mu$m wide ($\equiv  ~11 ~\kms$ and $1\farcs9$ ). Integration times were 1800 s in all cases and the spectra were calibrated in wavelength to $\pm 1 ~\kms$ accuracy when converted to heliocentric radial velocity (V$_{hel}$) against the spectrum of a Th/Ar arc lamp. Direct images of the jets were obtained also with MES in its imaging mode through the same H$\alpha$ + [N~ II] filter. For both the northern and southern jets the integration times were 1800 s.

Figure 1 shows a DSS image of NGC 1360 where the slit positions have been indicated and labeled. Insets in this Figure show the deep MES images of the jets, revealing its knotty and highly collimated structure. The spectra corresponding to slit a, that crosses the main body of the nebula in the N -- S direction is presented in Figure 2. The left panel in this Figure shows the full spectral range of the echelle order. The H$\alpha$ emission line is moderately split near the center and the line profile shows a continuous tilt from one extreme to the other. No [N~II] line emission is detected in the region covered by the slit, except for an isolated knot near the upper (north) end of the slit, this knot is blue-shifted with respect to the emission coming from the shell.  The right panel in this Figure shows the position -- velocity array for the region corresponding to H$\alpha$ were the characteristics mentioned above can be clearly appreciated. The velocity splitting (peak to peak) at the center of the profile amounts to 55.52 \kms with the approaching component at 22.9 \kms and the receding component at 75.42 \kms. This yields a 26.26 \kms expansion velocity and the systemic heliocentric velocity, as measured from the mid point between this line splitting is 46.85 \kms or 49.16 \kms if the average of the components is adopted.
The tilt in the line profiles runs from 33.41 \kms in the south to 78.21 \kms in the north. 

The four panels in Figure 3 show the spectra from the other four slits. The upper panels show the spectra from slits  b \& c for  the northern portion of the nebula, including the jets and the lower panels the spectra from slits d and e corresponding to the southern part of the nebula and jets. These spectra reveal that in all instances NGC 1360 behaves as a closed shell with closing, single line profiles at its border and the presence of a diffuse emission or halo extending from the main shell to the location were the jets start. The jets are distinctly displaced in velocity from the shell and halo, they are redshifted in the north and blue-shifted in the south by 29.94 \kms and -32.41 \kms, respectively. The  position -- velocity arrays for the jets are shown in detail in Figure 4 where the knots have been labeled. Table 1 lists for slits b, c and d the distance of each knot from the central star and their corresponding heliocentric velocity. The knots are seen to increase in velocity with distance from the central star, with heliocentric radial velocities reaching 123.6 \kms in the north (slit c) and -31.6 \kms in the south (slit d) corresponding to deprojected expansion velocities $\gtrsim$ 150 \kms, assuming an ellipsoidal shell 
inclined 30$^\circ$ with respect to the plane of the sky (see below).

\section {Discussion}

\citet{gol04} obtained a single line of  3 long-slits along the major axis (PA 27$^\circ$) of NGC 1360, reaching the edge of the northern jet. Our data provides a better and deeper  coverage of the bipolar jets and crosses the main nebula with a  N -- S slit. We find a very good agreement between our kinematic data described in the previous section and the results presented by \citet{gol04}. Furthermore, the faint outer region or halo that they describe in their H$\alpha$ image is clearly detected in our deep spectra from slits b, c, d \& e (see Figure 3). In view of these coincidences we adopt their model for NGC 1360 as a prolate ellipsoidal shell inclined 30$^\circ$ with respect to the plane of the sky as a reasonable representation of the general morphology of the nebula. 

The lack of apparent inner structure in NGC 1360 denotes the lack of a fast stellar wind from the central star at this stage. Since the kinematic age for the nebular shell is of the order of ten thousand years, this is not unexpected. We have retrieved IUE archival spectra (swp 06619, 18044, 16467 \& 55902) of the central star of NGC 1360 to search for P Cygni-type profiles in the usual N V $\lambda$1238, O V $\lambda$ 1371 and C IV $\lambda$ 1548 \AA  ~resonance lines. The IUE spectra confirm the absence of these line profiles that could be associated with a stellar wind. NGC 1360 fits into the expected structure described by  \citet{gar06} in situations like this. When the stellar wind becomes negligible, the hot, shocked bubble loses pressure and the thermal pressure of the photoionized region at the inner edge of the swept-up shell becomes dominant while photoionized material expands back toward the central star as a rarefaction wave, thus the photoionized gas fills the inner cavity.  The structure of NGC 1360 fits this description rather well. Furthermore, the components of the H$\alpha$ line profile from Figure 1 (slit a) are broader (35 -- 37 \kms) than that expected alone from  thermal width  ($\approx 22$ \kms), these line widths can be understood in terms of the readjustment of material during the back fill process. 

 NGC 1360 presents us with a rather peculiar situation: it shows no signs of having a binary system, it has a bipolar jet system, the fast stellar wind has faded and  the central star of NGC 1360 has a substantial magnetic field. Under these circumstances we have explored the case of  magnetic collimation during the development of this planetary nebula to try to understand its structural and kinematic evolution, including the history of the bipolar collimated outflows. We note, however, that the current lack of evidence for the presence of a binary nucleus does not preclude it from having existed in earlier stages, e.g. during the AGB phase and having merged after a common envelope interaction. In any case, the current rotational velocity of the central star (see next section) is quite moderate, as expected for a single white dwarf and does not provide clues as to the this possible early binary interaction. The MHD model is presented in the following section.

\section{Stellar Inputs and Hydrodynamical Simulation}

The simulation presented here has been performed with the hydrodynamical code 
ZEUS-3D (version 3.4) \citep{sto92, cla96}  
and details about the set up can be found in \citet{gar99}  
and \citet{gar05}  for the self-expanding grid technique.

We perform the two-dimensional simulation in spherical polar 
coordinates ($r,\theta$), with reflecting boundary conditions at the 
equator and the polar axis, and rotational symmetry assumed with respect 
to the latter.  Our grid consist of $200 \times 180$ equidistant zones in 
$r$ and $\theta$ respectively, with a radial extent of 0.1 pc (initially),
and an angular extent of $90^{\circ}$.  The innermost
radial zone lies at $r=2.5 \times 10^{-3}\,$pc from the central star.

The central star of NGC 1360 has been reported to have a surface magnetic 
field $B = -1343, 1708, 2832, 194 $ gauss \citep{jor05}.
The changes in sign and intensity have been atributed 
to the stellar rotation, with a $P > 0.75$ days, $v_{\rm rot} < 20  \kms $, 
with a stellar radius $ R = 0.3  \Ro $. Since Jordan et al. (2005) list a main sequence mass
for the central star of NGC 1360 of nearly 3 M$_\odot$, inferred from \citet{wei00} 
initial- final mass relation and the evidence indicates  that the  fast wind has died away,  
we have adopted the wind history corresponding to the stellar model
of $M_{\rm ZAMS} = 3.5 \Mo $ (Vassiliadis \& Wood 1994; and see Fig. 2 in 
Villaver et al. 2002). This particular model ceases its fast wind 
at 1,000 yr after the onset, allowing enough time to
produce a back-filling of its central cavity at $\approx$ 10,000 yr, \citep{gar06} which is  approximately
the observed kinematical age of NGC 1360.
The number of ionizing photons are computed from the same stellar 
evolution model (see Fig. 4 in Villaver et al. 2002).

Considering that there is no information on the evolution of the surface magnetic 
field, we have adopted for simplicity  a conservative value for 
$\sigma = 0.01$ at all times,
being
\beq
\sigma = \frac{B^2}{4 \pi \rho v_{\infty}^2} = 
\frac{B_{\rm s}^2 R_{\rm s}^2}{\Mdot v_{\infty} } \left(
\frac{ v_{\rm rot}}{v_{\infty}} \right)^2 
\eeq
the ratio of the magnetic field energy density to
the
kinetic energy density in the fast wind \citep{beg92, chev94}.

Figure 5 shows  four snapshots of the gas density  at 1,000, 3,000, 5,000 and
7,000 yrs after the onset of the fast wind. 
The first (top) panel  in Figure 5 shows
 the fast, free expanding wind region that surrounds the central star (dark grey), 
the terminal or reverse shock, the hot, 
shocked bubble (grey), the swept-up shell (white) with an outer, forward shock, and the 
slow wind (clear grey). This is the classical view of a two-wind model with an embeded 
toroidal magnetic field. Note that the inner shock is not spherical,
since the magnetic effects are larger at the polar axis, pushing the
shocked gas towards the star up to the point where it is balanced by
the ram pressure of the fast wind (the stagnation point). 
The rest of the panels show how the hot shocked bubble or the cavity, 
collapses after $\sim$ 2,000 yr, once the fast wind 
is switched off,  and how the ``back-filling'' of the 
photoionized gas finally fills up the available space.
During the back-filling, internal shocks close to the axis are formed due to
the collision of opposite, incoming streams of gas. The bouncing of the gas,
produces an inner bipolar like outflow at the center of the nebula, visible at
the lower panel of Figure 5.
The radial spikes in the AGB slow wind are due to transients in the trapping of the 
ionization front.
The model is marginally optically thin, producing spikes whenever the conditions
turn into optically thick conditions \citep{gar06}.

The current characteristics of NGC 1360 are best reproduced when the model reaches
11,000 yrs, as shown in Figure 6 for the gas density, radial velocity and emission measure 
of the gas when considering a 30$^\circ$ \citep{gol04}
tilt from the plane of the sky for the nebula.

\citet{gol04} derive a simple kinematic age of $\approx$ 5000 years for the jets of NGC 1360 which  implies that the jets were launched with constant velocity when the nebula was already well developed at half its current kinematic age and by a time when the fast stellar wind was probably starting to weaken. This would indeed be a rare case since jet formation is expected to occur during the very early stages of development of the planetary nebula,  right at the end of the AGB stage.
To trace the evolution of the model in velocity, 1-dimensional plots of radial velocity {\it vs} distance from the central star along the polar axis are shown in Figure 7 for time intervals of $10^3$ yrs. 
The velocities and distances for the individual knots in the northern jet that are intersected by the slit "c" have been drawn from Table 1, and are plotted in this figure as solid squares,  (notice that the velocities plotted are V$_{hel}$ - V$_{sys}$) showing a very good fit for the model at $\sim 11,500 $ yrs. Again, these velocity plots
take into account a  $30^{\circ}$ tilt with respect to the plane of the sky. The plot provide a good explanation of the current state of the jets, their origin as launched by the magnetized stellar wind, their current location and observed increasing outward velocities and conciliate their age with that of the nebula.

\section{Conclusions}

 NGC 1360 is a large planetary nebula with a mild expansion velocity, broad line components and a system of bipolar, fast, collimated outflows that increase in velocity with distance. The fast stellar wind from the central star has died away at least a few thousand years ago and a back filling process has modified its structure producing a smooth, nearly featureless and elongated high excitation nebula. Past attempts to identify a binary core in this object have yielded negative results. The central star conserves a strong magnetic field  indicating that a magnetized stellar wind could have been responsible for axial collimation in the development of this planetary nebula. We have therefore studied the history of this nebula considering a process of magnetic collimation and then switching off the stellar wind and letting it evolve to its present time. The resulting model is able to successfully reproduce many of the observed key features in NGC 1360.

\acknowledgments

MTG-D gratefully acknowledeges the support of a postdoctoral grant from CONACYT.
This research has benefited from the finantial support of DGAPA-UNAM through grants
IN116908, IN108506-2 \& IN108406-2.

\clearpage

\begin{table}[htbp]
\caption{}
\begin{center}
\begin{tabular}{ccc}\hline
\hline
 & Position & V$_{\small{{\textrm {hel}}}}$ \\
Label & (arcsec) & (\kms) \\\tableline
\multicolumn{3}{c}{Slit b}\\\hline
A & 264.8 & 106.45 \\
B & 277.0 & 108.45 \\
C & 282.0 & 108.95 \\\hline
\multicolumn{3}{c}{Slit c}\\\hline
A & 285.8 & 116.7 \\
B & 295.8 & 118.9 \\
C & 302.0 & 119.5 \\
D & 315.9 & 121.8 \\
E & 329.3 & 123.6 \\\hline
\multicolumn{3}{c}{Slit d}\\\hline
A & 241.0 & -16.0 \\
B & 246.8 & -24.6 \\
C & 267.4 & -31.3 \\
D & 289.9 & -31.6 \\ 
E & 301.1 & -31.6 \\
\hline
\end{tabular}
\end{center}
\end{table}

\clearpage



\begin{figure}
\plotone{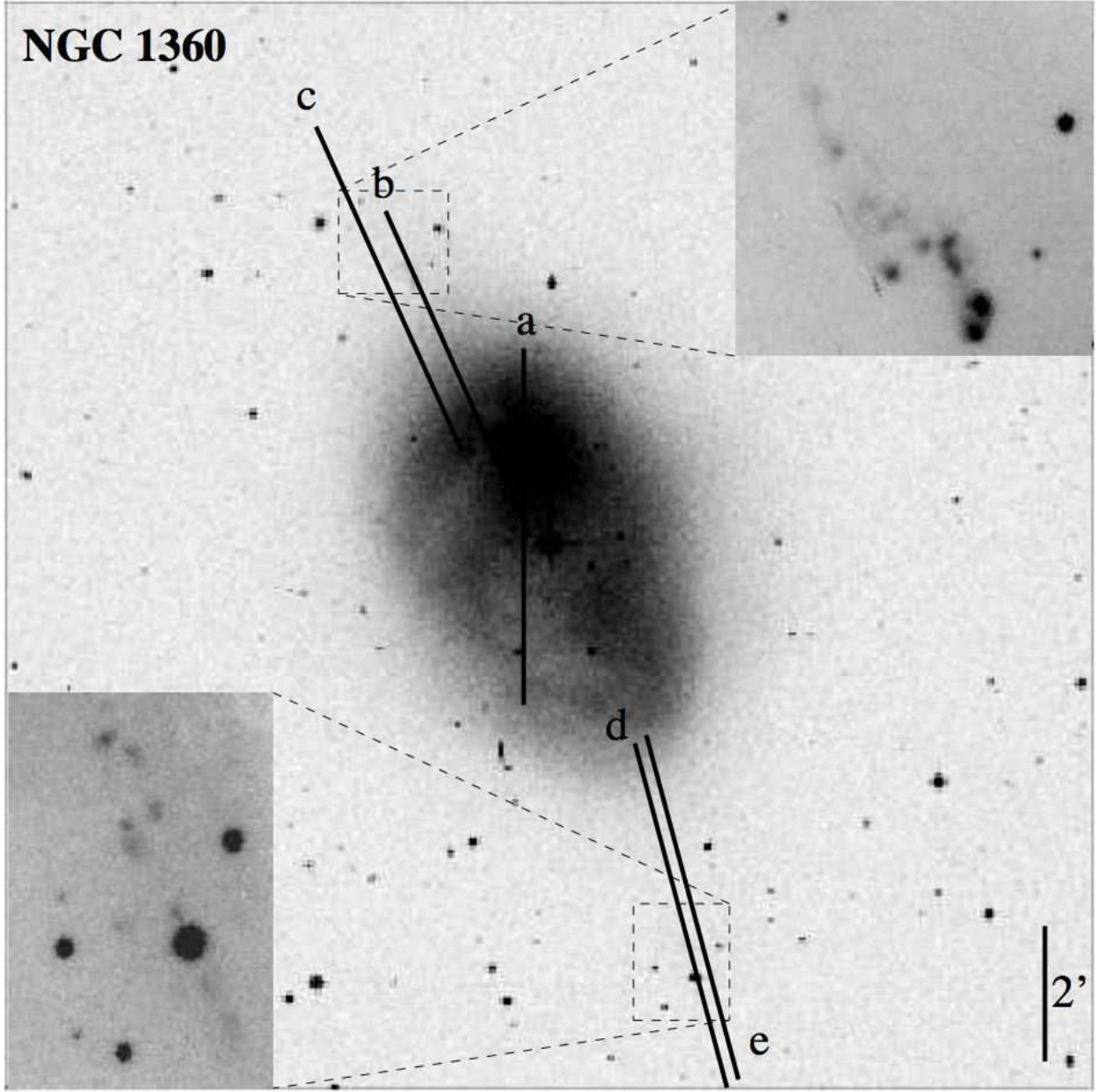}
\caption{Image of NGC 1360 from the DSS. Insets are deep images from MES-SPM that show enlargements of the bipolar collimated outflows. The location of the slits are indicated and labeled on the image.}
\end{figure}

\clearpage

\begin{figure}
\plotone{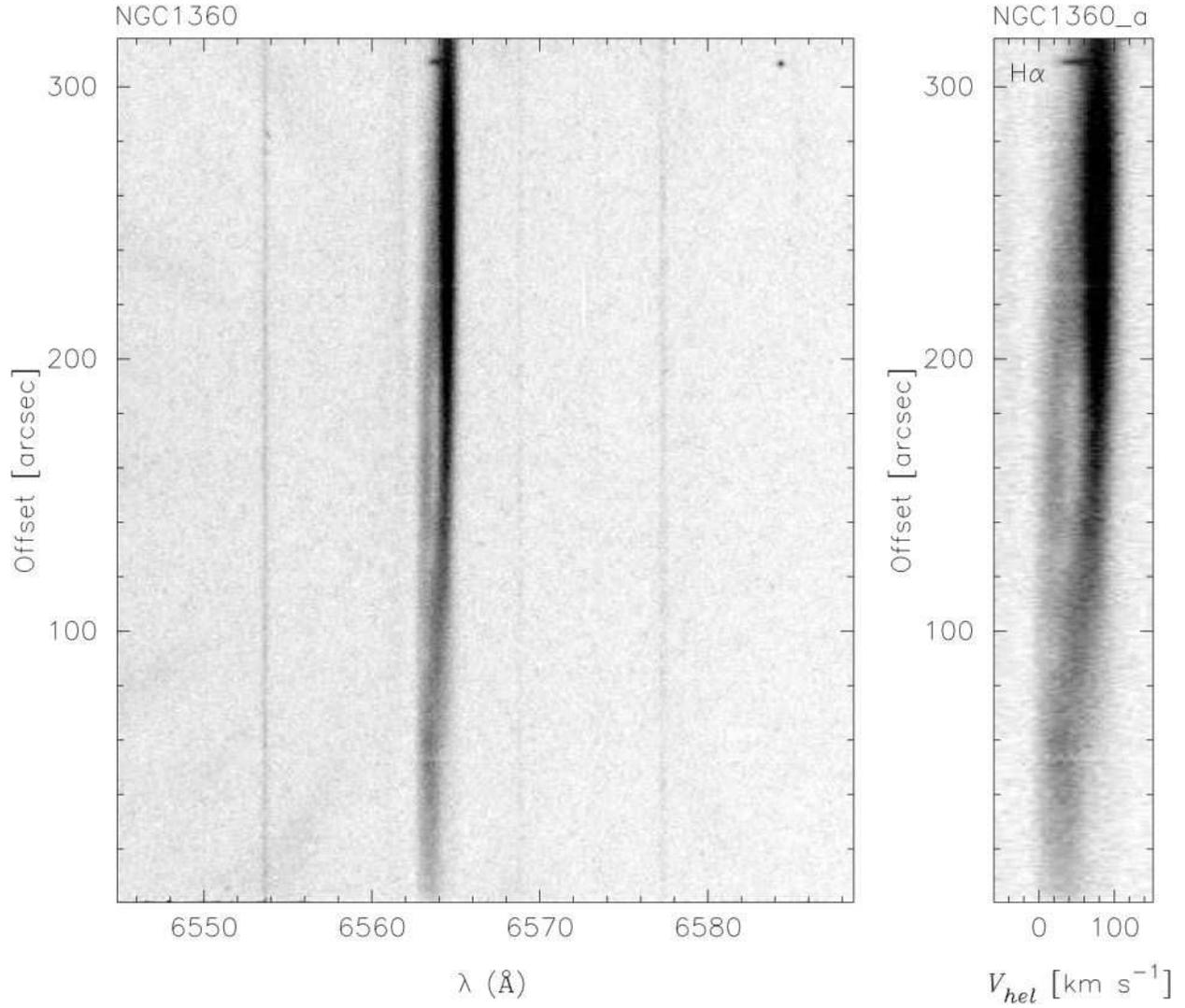}
\caption{The long-slit spectra from slit a showing the full spectral range of the echelle order. Only the H$\alpha$ emission line is detected, split, over the main body of NGC 1360 , except for an isolated knot in the northern edge (top) that also emits in [N~II]. Right panel, an enlarged, wavelength calibrated subsection of the spectrum covering H$\alpha$.}
\end{figure}

\clearpage

\begin{figure}
\plotone{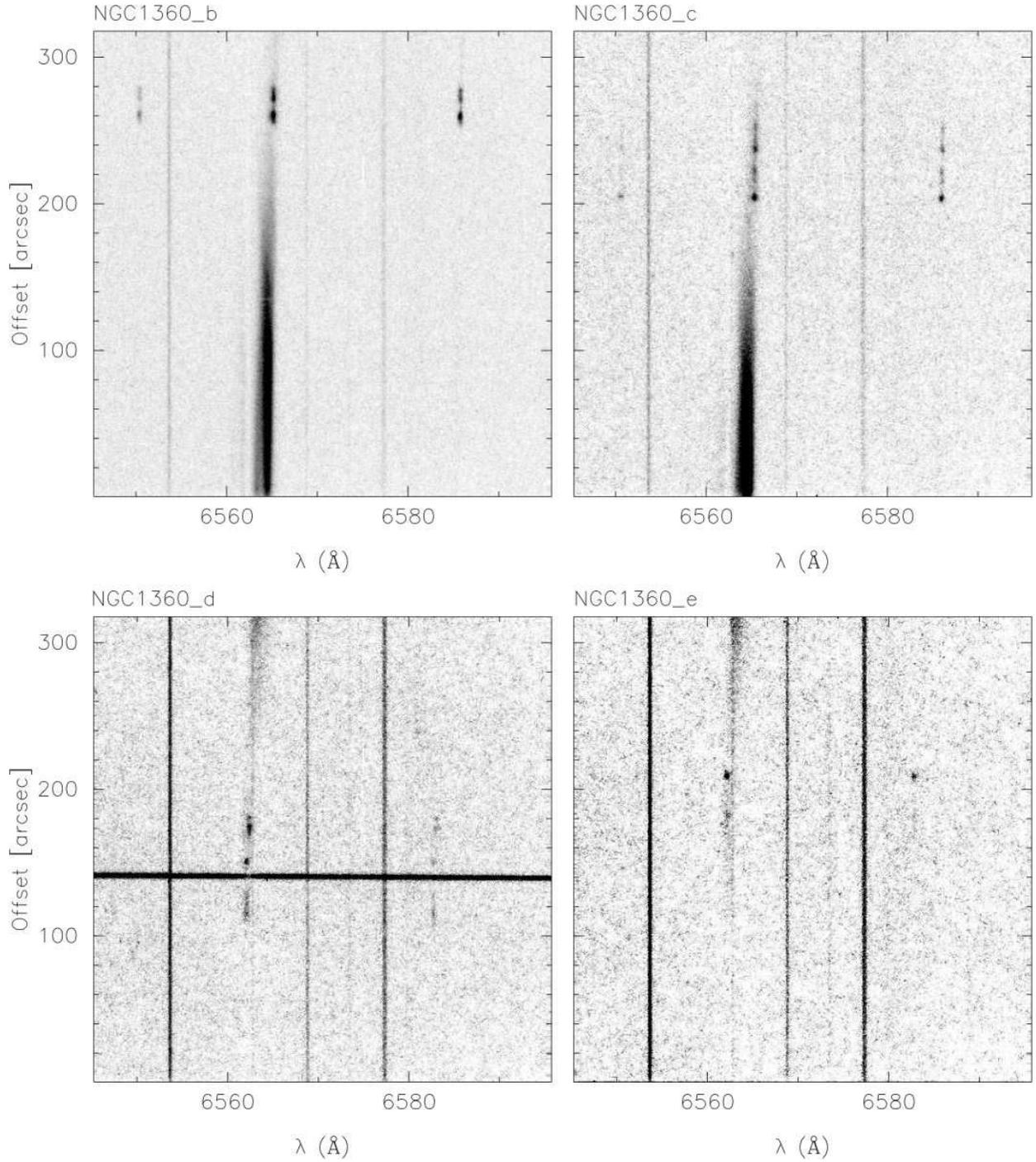}
\caption{The long-slit spectra from positions b, c, d \& e, showing the full spectral range of the wavelength calibrated echelle order. The northern jets are located at the top of slits b and c and the southern one at the bottom of  slits d and e.}
\end{figure}

\clearpage

\begin{figure}
\epsscale{.70}
\plotone{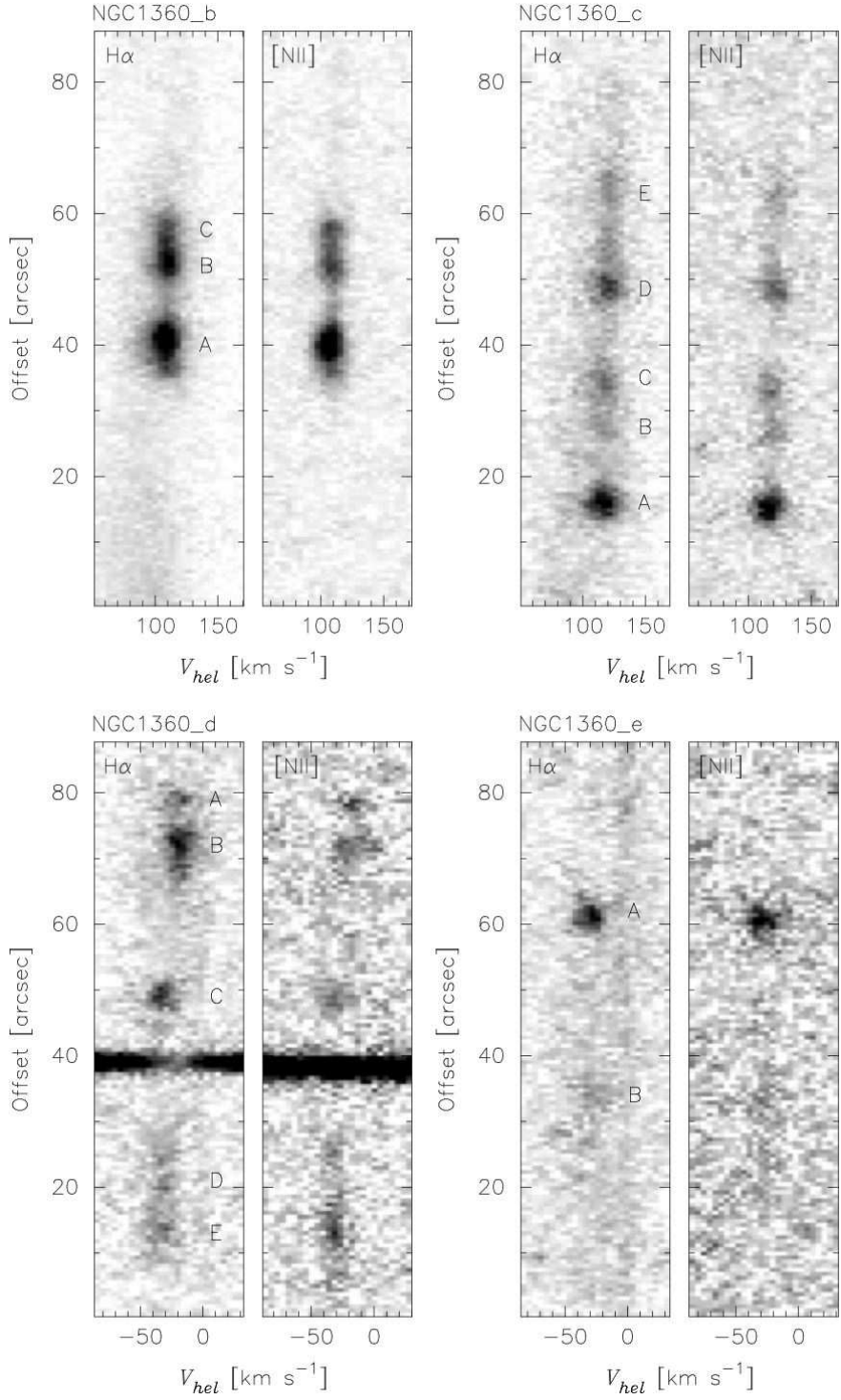}
\caption{Position -- Velocity diagrams of the northern and southern jets. Individual knots are labeled.}
\end{figure}

\clearpage

\begin{figure}
\epsscale{.99}
\plotone{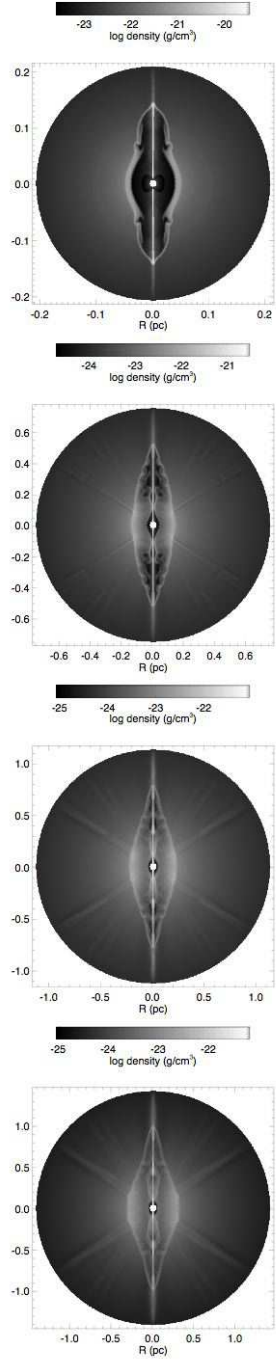}
\caption{Snapshots of gas density at 1000, 3000, 5000 and 7000 yrs (from the top) after the onset of the fast wind.}
\end{figure}

\clearpage

\begin{figure}
\epsscale{.99}
\plotone{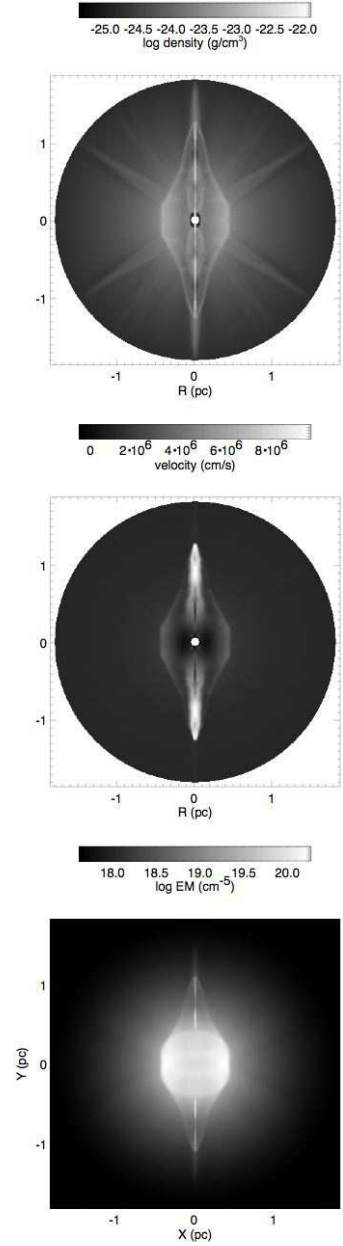}
\caption{Snapshot of the model at 11,000 yrs. Top panel: gas density; middle panel: radial velocity; bottom panel: emission measure of the gas, as projected with a $30^\circ$ tilt from the plane of the sky.}
\end{figure}

\clearpage

\begin{figure}
\plotone{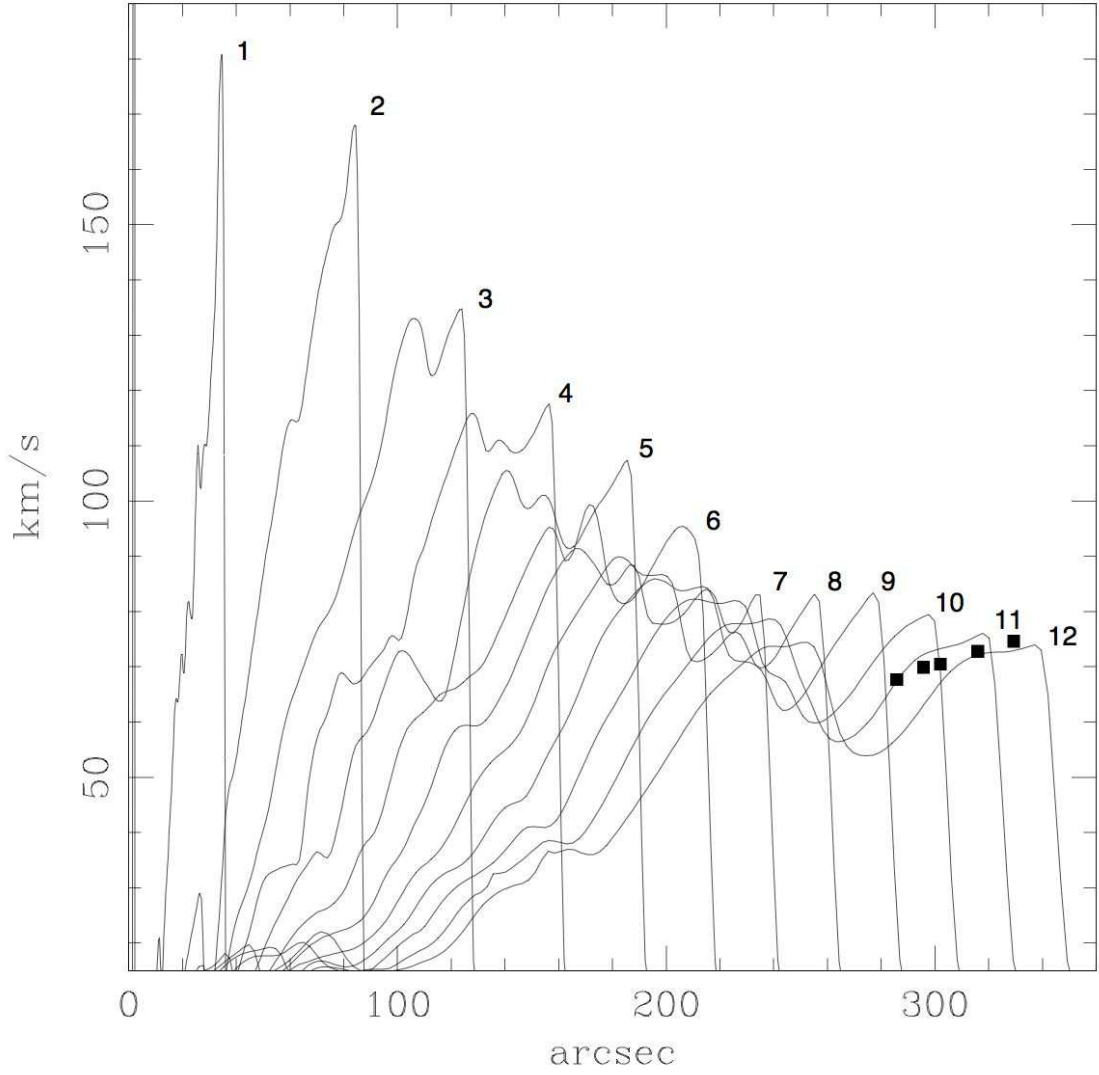}
\caption{Velocity {\it vs} distance from the central star plots for the model along the polar axis for  time intervals in units of 10$^3$ yrs. The individual northern knots from slit c are shown as solid squares, V$_{sys}$ has been subtracted to the V$_{hel}$ listed in Table 1}
\end{figure}

\end{document}